\documentclass[structabstract]{aa}  

\usepackage{graphicx,psfig}
\usepackage{txfonts}

\begin{document}

\title{OGLE2-TR-L9b: An exoplanet transiting a fast-rotating F3 star\footnote{Based on observations collected at the European Organisation for Astronomical Research in the Southern Hemisphere, Chile (280.C-5036(A))}}

\authorrunning{Snellen et al.}
\titlerunning{An extrasolar planet transiting a fast-rotating F3 star}

\author{Snellen I.A.G.\inst{1}, Koppenhoefer J.\inst{2,3}, van der Burg R.F.J.\inst{1}, Dreizler S.\inst{4}, Greiner J.\inst{3}, de Hoon M.D.J.\inst{1}, Husser T.O.\inst{5}, Kr\"uhler~T.\inst{3,6}, Saglia R.P.\inst{3}, Vuijsje F.N.\inst{1}}

\institute{
 Leiden Observatory, Leiden University, Postbus 9513, 2300 RA, Leiden, The Netherlands \and
 Universit\"ats-Sternwarte M\"unchen, Munich, Germany  
\and
Max-Planck-Institut f\"ur extraterrestrische Physik, Giessenbachstrasse, D-85748 Garching, Germany 
\and
Institut f\"ur Astrophysik, Georg-August-Universit\"at G\"ottingen, Friedrich-Hund-Platz 1, 37077 G\"ottingen, Germany
\and
South African Astronomical Observatory, P.O. Box 9, Observatory 7935, South Africa
\and Universe Cluster, Technische Universit\"at M\"unchen, Boltzmannstra\ss e 2, D-85748, Garching, Germany}

\offprints{snellen@strw.leidenuniv.nl}

\date{}

\abstract{Photometric observations for the OGLE-II microlens monitoring 
campaign have been taken in the period 1997$-$2000. All light curves of this campaign have recently been made public. Our
analysis of these data has revealed 13 low-amplitude transiting objects among 
$\sim$15700 stars in three Carina fields towards the galactic disk. One of 
these objects, OGLE2-TR-L9 (P$\sim$2.5 days), turned out to be an excellent 
transiting planet candidate.}
{In this paper we report on our investigation of the true nature of OGLE2-TR-L9,
by re-observing the photometric transit with the aim to determine the transit 
parameters at high precision, and by spectroscopic observations, to estimate the properties 
of the host star, and to determine the mass of the transiting object through
radial velocity measurements.}
{High precision photometric observations have been obtained in $g'$, $r'$, $i'$, 
and $z'$ band simultaneously, using the new GROND detector, mounted on 
the MPI/ESO 2.2m telescope at La Silla. Eight epochs of high-dispersion 
spectroscopic 
observations were obtained using the fiber-fed FLAMES/UVES Echelle 
spectrograph, mounted on ESO's Very Large Telescope at Paranal.} 
{The photometric transit, now more than 7 years after the 
last OGLE-II observations, was re-discovered only $\sim$8 minutes from its 
predicted time. The primary object is a fast rotating F3 star, with
$v$sin$i$=39.33$\pm$0.38 km/s, T=6933$\pm$58 K, log g = 4.25$\pm$0.01, and 
[Fe/H] = $-$0.05$\pm$0.20. The transiting object is an extrasolar planet
with M$_{\rm{p}}$=4.5$\pm$1.5 M$_{\rm{Jup}}$ and 
R$_{\rm{p}}$=1.61$\pm$0.04R$_{\rm{Jup}}$. Since this is the first planet 
found to orbit a fast rotating star, the uncertainties
in the radial velocity measurements and in the planetary mass are larger than 
for most other planets discovered to date. The rejection of possible blend scenarios was based on a quantitative analysis of the multi-color photometric data. A stellar blend scenario 
of an early F star with a faint eclipsing binary system is excluded, due to the 
combination of 
1) the consistency between the spectroscopic parameters of the 
star and the mean density of the transited object as determined from 
the photometry, and 2) the excellent agreement between the transit signal 
as observed at four different wavelengths.}{}


   \maketitle

\section{Introduction}
Transiting extrasolar planets allow direct 
measurements of their fundamental parameters, such as planet mass, 
radius, and mean density. Furthermore, their atmospheres can be 
probed through secondary eclipse observations (e.g. Charbonneau et al. 
2005; Deming et al. 2005; Knutson et al. 2007), and atmospheric transmission 
spectroscopy (e.g. Charbonneau et al. 2002; Tinetti et al. 2007; 
Snellen et al. 2008). This makes transiting exoplanets of great scientific
value.

Many photometric monitoring surveys are currently underway. Several 
of these surveys are very successful, such as 
the transit campaigns of the Optical Gravitational Lens Experiment 
(OGLE-III; 7 planets; e.g. Udalski et al. 2008), 
the Trans-Atlantic Exoplanet Survey (TrES; 4 planets; e.g. 
Mandushev et al. 2007 ),
the Hungarian Automated Telescope Network (HATNet; 9 planets; e.g. 
Shporer et al. 2008), the XO survey 
(5 planets; e.g. Burke et al. 2007), and the Wide Area Search for 
Planets (SuperWASP; 15 planets; e.g. Anderson et al. 2008). In addition, 
CoRoT is targeting planet transits from space (4 planets; e.g. 
Aigrain et al. 2008), ultimately aimed at finding Earth or super-Earth size 
planets. 

In this paper we present a new transiting extrasolar planet, OGLE2-TR-L9\,b.
The system has an I-band magnitude of I=13.97, and an orbital period of 
$\sim$2.5 days. The host star is the fastest rotating and hottest (main sequence) star around
which an orbiting extrasolar planet has been detected to date.
The transit system was the prime planetary candidate from a sample 
of thirteen objects presented by Snellen et al. (2007; S07). They were drawn 
from the online database of the second phase of the OGLE project, 
a campaign primarily aimed at finding microlensing events, 
conducted between 1997 and 2000 (OGLE-II; Szymanski 2005; Udalski et al. 1997).
The low amplitude transits were discovered among the light curves of 
$\sim$15700 stars, with 13.0$<$I$<$16.0, located in three Carina fields 
towards the galactic plane. Note that the light curve data have a 
different cadence than usual for transit surveys, with 1$-$2 photometric
points taken per night, totalling 500$-$600 epochs over 4 years.
This work shows that such a data set is indeed sensitive to 
low-amplitude transits, and can yield transiting extrasolar planets.

In section 2 of this paper we present the analysis of new transit photometry of 
OGLE2-TR-L9, taken with the GROND instrument mounted on the MPI/ESO 2.2m
telescope at La Silla. The particular goals of these observations are 
the re-discovery of the transit, improved transit parameters and 
orbital ephemeris. 
In section 3 the spectroscopic observations with the FLAMES/UVES multi-fiber
spectrograph on ESO's Very Large Telescope are described, including a 
description of the analysis of the radial velocity variations, of the bisector 
span, and the spectroscopic parameters of the host star.
In section 4 and 5 the stellar and planetary parameters with their 
uncertainties are determined, and arguments against a stellar blend scenario
laid out. The results are discussed in section 6.

\section{Transit photometry with GROND}
\begin{figure}
\psfig{figure=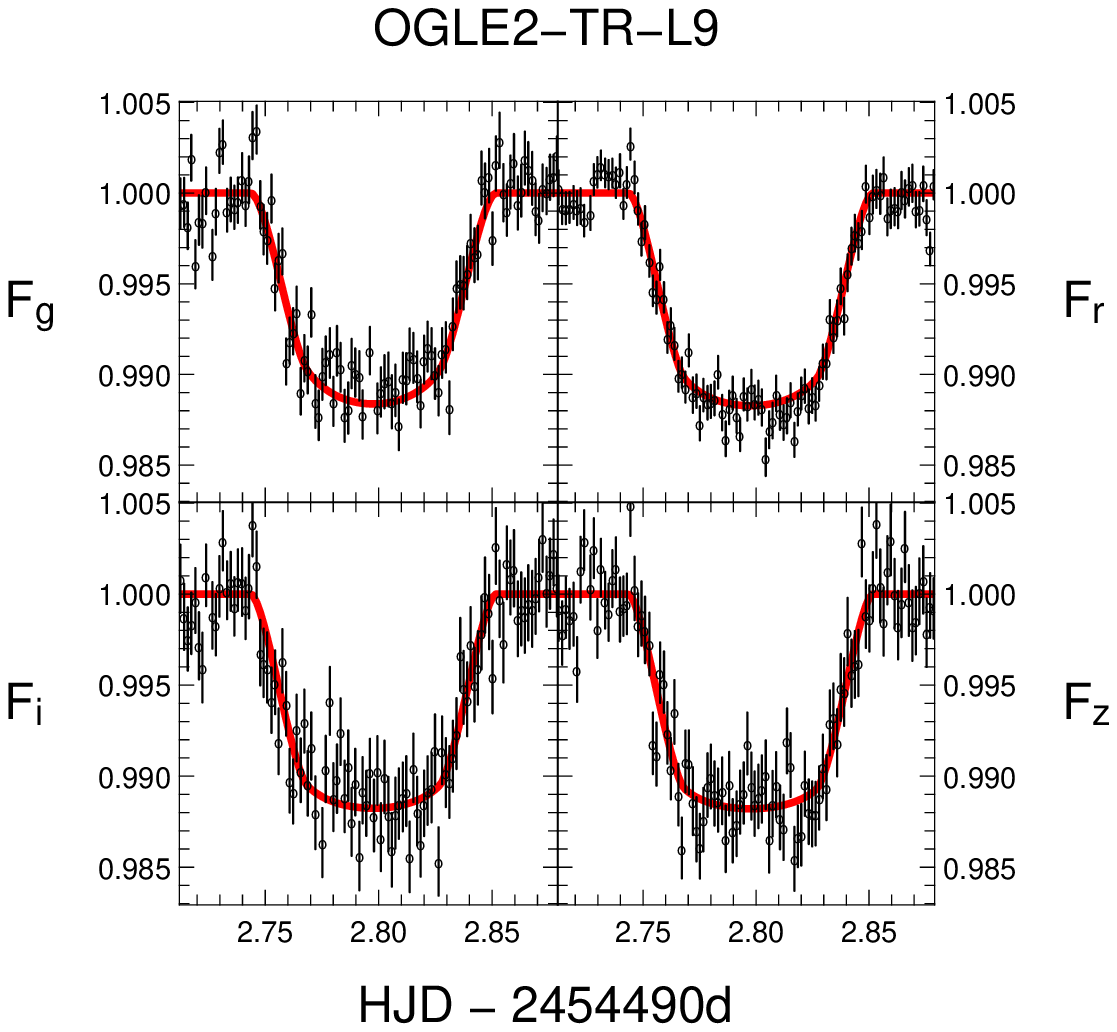,width=0.5\textwidth}
\caption{\label{trans} 
Transit lightcurves of OGLE-TR-L9 in g$'$, r$'$, i$'$ and z$'$ observed simultaneously with the GROND instrument, mounted on the MPI/ESO 2.2m telescope. The line shows the best model fit for the combined light curves, as discussed in the text. 
 }
\end{figure}

\begin{table}
  \caption{Limb-darkening coefficients used for the transit fitting, taken 
from Claret (2004) for a star with metallicity $[Fe/H]$=0.0,
surface gravity \mbox{log $g$}=4.5, and effective temperature
$T_{eff}$=7000K.}
\centering
  \begin{tabular}{ccc} 
  filter  & $\gamma_1$ & $\gamma_2$  \\ \hline
  g$'$      & 0.3395     & 0.3772      \\ 
  r$'$      & 0.2071     & 0.3956      \\ 
  i$'$      & 0.1421     & 0.3792      \\ 
  z$'$      & 0.0934     & 0.3682      \\ 
  \end{tabular}
  \label{limb} 
\end{table}

\subsection{Data acquisition and analysis}
We observed one full transit of OGLE-TR-L9 with GROND (Greiner et al. 2008), 
which is a gamma ray burst follow-up instrument mounted on the MPI/ESO 2.2m 
telescope at the La Silla observatory. GROND
is a 7-channel imager that allows to take 4 optical (g$'$r$'$i$'$z$'$) and 3 near
infrared (JHK) exposures simultaneously.
On  January 27, 2008, a
total of 104 images in each optical band and 1248 images in each near
infrared band were taken. The JHK-images turned out to have
insufficient signal to noise to detect the transit, and will not be
considered further. Using exposure times of 66 seconds and a cycle rate of 
2.5 minutes, we covered a period of about 4 hours centered on the predicted 
transit time.
All optical images have been reduced with the {\it mupipe} software
developed at the University Observatory in 
Munich\footnote{http://www.usm.lmu.de/$\sim$arri/mupipe/}. After the initial
bias and flatfield corrections, cosmic rays and bad pixels were masked
and the images resampled to a common grid. The frames did not suffer from
detectable fringing, even in z-band. Aperture photometry was performed on 
OGLE2-TR-L9 and eight to ten interactively selected reference stars, after 
which light curves were created for each of the 4 bands.
The aperture radius was chosen to be 12 pixels, corresponding to 1.9 
arcseconds, with a seeing of typically 1.1 arcseconds during the observations. 
The sky was determined in an annulus between 20 and 30 pixels from the object 
positions. The rms in the individual light curves of the reference stars was 
in all cases better than 0.3\%. This resulted in typical precisions 
of the relative fluxes better than 0.2\%.

\subsection{Fitting the transit light curves}

The lightcurves in g$'$, r$'$, i$'$, and z$'$, were fitted with analytic
models as given by Mandel \& Agol (2002).  We
use quadratic limb-darkening coefficients taken from
Claret (2004), for a star with metallicity $[\rm{Fe/H}]$=0.0,
surface gravity \mbox{log $g$}=4.5, and effective temperature
$T_{\rm{eff}}$=7000K (very close to the spectroscopic parameters of 
the star as determined below). The values of the limb-darkening 
coefficients are given in Table \ref{limb}.
Using a simultaneous fit to  all 4 lightcurves we derived the mean
stellar density, $M_{\rm{star}}$ / $R^3_{\rm{star}}$ in solar units, the radius 
ratio $R_{\rm{planet}}$ / $R_{\rm{star}}$, the impact parameter $\beta_{\rm{impact}}$ in units of
$R_{\rm{star}}$, and the timing of the central transit. Together with a scaling 
factor for each band, there were eight free parameters to fit.

The light curves and the model fits are shown in
Fig. 1., and the resulting parameters are listed in
Table 2. All lightcurves fit well to the model except
for the g-band lightcurve, which is attributed to the significantly more noisy 
light curve, and the poorly determined baseline, particularly before ingress.

\section{Spectroscopic Observations with UVES/FLAMES}

We observed OGLE2-TR-L9 with the 
UV-Visual Echelle Spectrograph (UVES; Dekker et al. 2000), mounted
at the Nasmyth B focus of UT2 of ESO's Very Large 
Telescope (VLT) at Paranal, Chile. The aims of these observations were
to estimate the spectroscopic parameters of the host star, and 
to determine the radial velocity variations.
The observations were performed in fiber mode, with UVES connected to the 
FLAMES fiber facility (Pasquini et al. 2002), with 7 science fibers and 
with simultaneous thorium-argon wavelength calibration ({\sl UVES7 mode}).
Apart from our main target, fibers were allocated to two other OGLE-II 
transit candidates from S07 (OGLE2-TR-L7 and OGLE2-TR-L12), and three 
random stars within the 25$'$ FLAMES field. In addition, one fiber was 
positioned on empty sky. A setup with a central wavelength of 580 nm 
was used, resulting in a wavelength coverage of 4785$-$6817$\AA$ over
two CCDs, at a resolving power of R=47000. Since the upper CCD turned
out to cover only a small number of strong stellar absorption lines,
in addition to suffering from significant telluric contamination,
only the lower CCD (4785$-$5729$\AA$) was used for further analysis.

Eight observations were taken in Director's Discretionary Time, in service 
mode in the course of  December 2007 and January 2008,
spread in such a way that the data would be evenly distributed in orbital phase
of our main target (see table 3).
The data were analysed using the {\sl midas}-based UVES/FLAMES pipeline 
as provided by ESO, which results in fully reduced, wavelength calibrated 
spectra. Since we were concerned about the wavelength calibration of the
fifth epoch (see below), we also analysed the data using purpose-build IDL 
routines. No significant differences in the wavelength solutions were found.
The resulting signal-to-noise per resolution element, in the central part of 
the orders, varies between $\sim$10 and 20  over the different epochs 
(see table 3).

\subsection{Determination of stellar spectroscopic parameters}

\begin{figure}
\psfig{figure=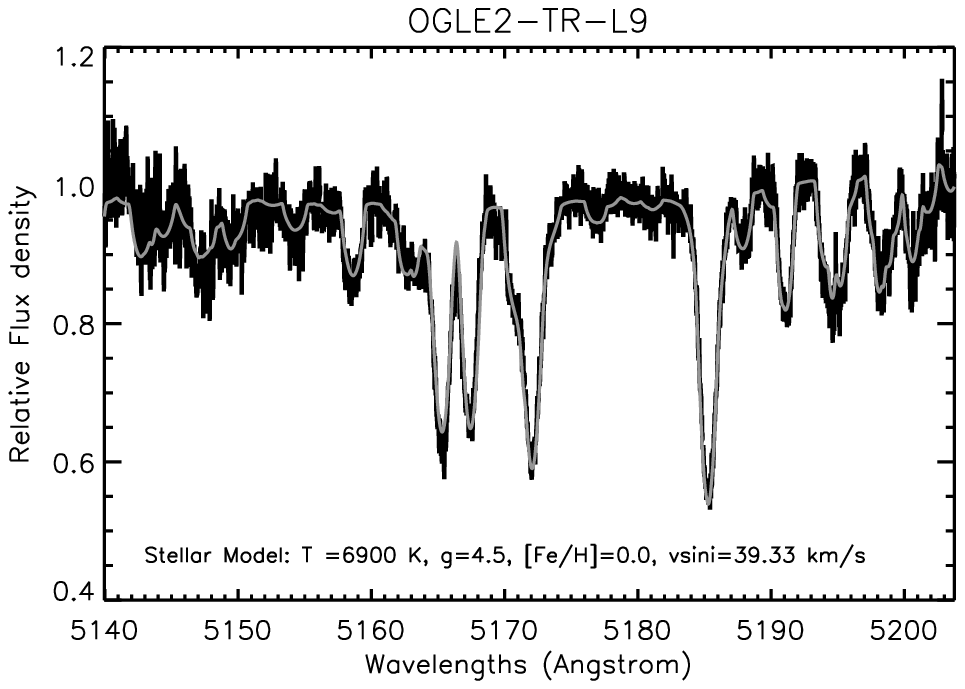,width=0.5\textwidth}
\caption{\label{spec} The central part of one order of the combined UVES 
spectrum of OGLE2-TR-L9,
with overplotted a synthetic spectrum with T=6900 K, g=4.5, [Fe/H]=0, and v{\sl sin}i = 39.33 km/sec.}
\end{figure}

\begin{table}
\caption{The transit, host star, and planetary companion parameters as 
determined from our photometric and spectroscopic observations.}
\begin{tabular}{rcl}\hline
\multicolumn{2}{l}{Transit:}&\\\hline
$\rho_s$ &=& 0.4260$\pm$0.0091 $\rho_{\rm{sun}}$\\ 
R$_p$/R$_s$ &=& 0.10847$\pm$0.00098\\
$\beta_{\rm{impact}}$ &=& 0.7699$\pm$0.0085\\
t$_0$               &=& 2\,454\,492.79765$\pm$0.00039 HJD \\ 
P &=& 2.4855335$\pm$7$\times10^{-7}$ d\\ \hline
\multicolumn{2}{l}{Host star:}&\\ \hline
Coord. (J2000) &=& 11$^{\rm{h}}$ 07$^{\rm{m}}$ 55.29$^{\rm{s}}$ $\ -$61$^\circ$ 08$'$ 46.3$''$ $\ \ \ $\\ 
I mag &=& 13.974\\ 
I$-$J mag &=& 0.466$\pm$0.032\\
J$-$K mag &=& 0.391$\pm$0.049\\
T &=& 6933$\pm$58 K\\
log g &=& 4.47$\pm$0.13$^{*}$\\
log g &=& 4.25$\pm$0.01$^{**}$\\
$[Fe/H]$ &=& $-$0.05$\pm$0.20\\
v{\sl sin}i &=& 39.33$\pm$0.38 km/s\\
R$_s$ &=&1.53$\pm$0.04 R$_{\rm{sun}}$ \\
M$_s$ &=&1.52$\pm$0.08 M$_{\rm{sun}}$ \\
Age &$<$&0.66 Gyr \\ \hline
\multicolumn{1}{l}{Planetary Companion:}&&\\ \hline
K&=&510$\pm$170 m/s\\
$i$ &=&79.8$\pm$0.3$^\circ$\\
$a$ &=&0.0308$\pm$0.0005 AU\\
R$_p$ &=& 1.61$\pm$0.04 R$_{\rm{jup}}$\\ 
M$_p$ &=& 4.5$\pm$1.5 M$_{\rm{jup}}$\\ \hline
\multicolumn{3}{l}{$^{*}$Determined from the spectroscopic analysis}\\
\multicolumn{3}{l}{$^{**}$Determined from mean stellar density combined with the}\\ 
\multicolumn{3}{l}{$\ \ \ $evolutionary tracks}\\
\end{tabular}
\end{table}

\begin{table}
\caption{Spectroscopic observations of OGLE2-TR-L9 taken with UVES/FLAMES.
The first three columns give the Heliocentric Julian Date, the planet's 
orbital phase at the time of observation, and the signal-to-noise ratio
of the spectra per resolution element in the center of the middle order. 
Column 4 and 5 give the radial velocity and the bisector span measurements.}
\begin{tabular}{ccrrr} \hline
HJD     &  Orbital  &  SNR & RV         &   BiS        \\
-2450000 &  Phase    &      & km s$^{-1}$ & km s$^{-1}$   \\ \hline

 4465.8421&0.157&  19.0&    1.090$\pm$0.224&     -0.076$\pm$0.482\\
 4466.8583&0.566&  16.6&    1.204$\pm$0.276&      0.173$\pm$0.380\\
 4468.7219&0.316&  11.2&    0.842$\pm$0.212&     -0.233$\pm$0.683\\
 4472.7447&0.934&  14.7&    1.105$\pm$0.231&     -0.492$\pm$0.460\\
 4489.6749&0.746&   9.0&    2.345$\pm$0.376&      0.123$\pm$1.101\\
 4490.8382&0.214&  15.3&    0.472$\pm$0.318&      0.235$\pm$0.339\\
 4493.7697&0.393&  19.9&    1.187$\pm$0.187&     -0.150$\pm$0.432\\
 4496.7703&0.601&  19.0&    1.190$\pm$0.272&      0.418$\pm$0.545\\ \hline

\end{tabular}

\end{table}

The spectroscopic parameters of the star, v{\sl sin}i, surface temperature,
surface gravity, and metallicity, were determined from the SNR-weighted,
radial velocity shifted combination of the eight epochs taken with UVES.
This combined spectrum has a signal-to-noise ratio of $\sim$42 in the 
central areas of the orders.
Detailed synthetic spectra were computed using the interactive data 
language (IDL) interface SYNPLOT (I. Hubeny, private communication) to the 
spectrum synthesis program SYNSPEC (Hubeny et al. 1995), utilising 
Kurucz model atmospheres\footnote{http://kurucz.harvard.edu/grids.html}. 
These were least-squares
fitted to each individual order of the combined UVES spectrum.
The final atmospheric parameters were taken as the average values
across the available orders. The uncertainties in the fitted parameters 
estimated using a $\chi$-square analysis and from the scatter between the 
orders, give similar results, of which the latter are adopted. 
  
The best fitting parameters and their uncertainties are given in table 2. 
One order of the combined UVES spectrum is shown in figure \ref{spec}, 
showing the 
Mg$_{\rm{b}}$ 5170$\AA$ complex, with the synthetic spectrum 
with T=6900 K, log g=4.5, [Fe/H]=0, and v{\sl sin}i = 39.33 km/sec, 
overplotted.

\subsection{Radial velocity measurements}

The orders of the eight spectra were first cosine-tapered to reduce
edge effects. Cross-correlations were performed using the best-fitted 
velocity-broadened synthetic spectrum, as determined above, as a reference. 
The spectrum of the sky-fiber indicated that the sky contribution 
was typically of the order of $\sim$0.5\%. However, for the observation at
0.74 orbital phase (epoch 5), the 
relative sky levels were an order of magnitude larger, 
 due to a combination of bad seeing and full moon.
We therefore subtracted the sky spectrum from all target spectra before 
cross-correlation. 

\begin{figure}
\psfig{figure=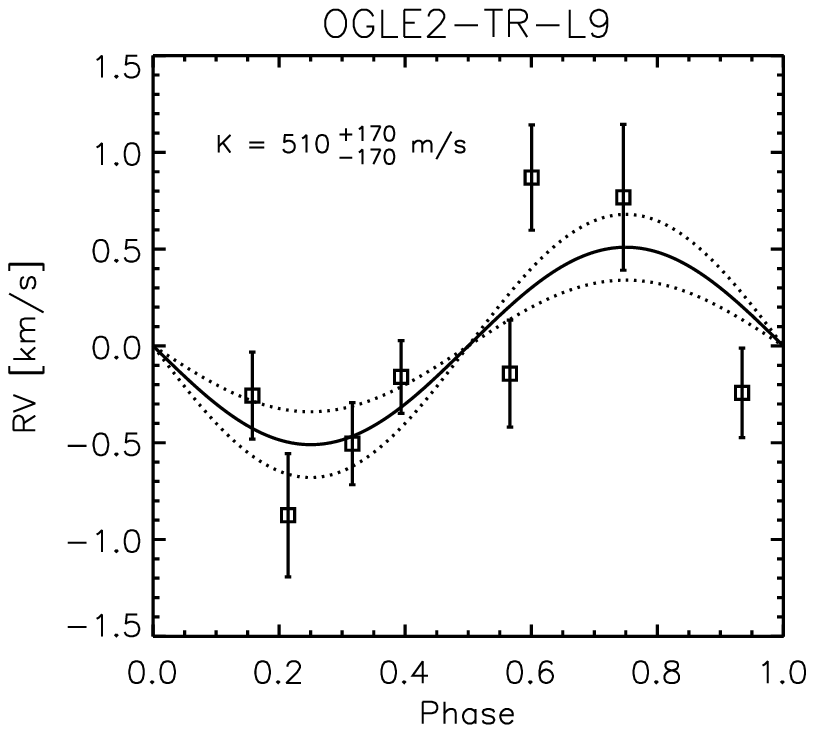,width=0.5\textwidth}
\caption{\label{rv} The radial velocity measurements of OGLE2-TR-L9 as function
of orbital phase from the ephemeris of the transit photometry.}
\end{figure}

The resulting radial velocity data are listed in table 3, corrected to
heliocentric values. The uncertainties are estimated from the variation 
of the radial velocity fits between the different orders. The final radial 
velocity data as function of orbital
phase (the latter determined from the transit photometry), 
are shown in figure \ref{rv}.
The data were fitted with a sine function with the
amplitude of the radial velocity variations, $K$, and a zero-point, V$_0$,
as free parameters. The radial velocity amplitude was determined at K$= 510\pm170$ m/s, with V$_0$=+0.2 km/s.

\begin{figure}
\psfig{figure=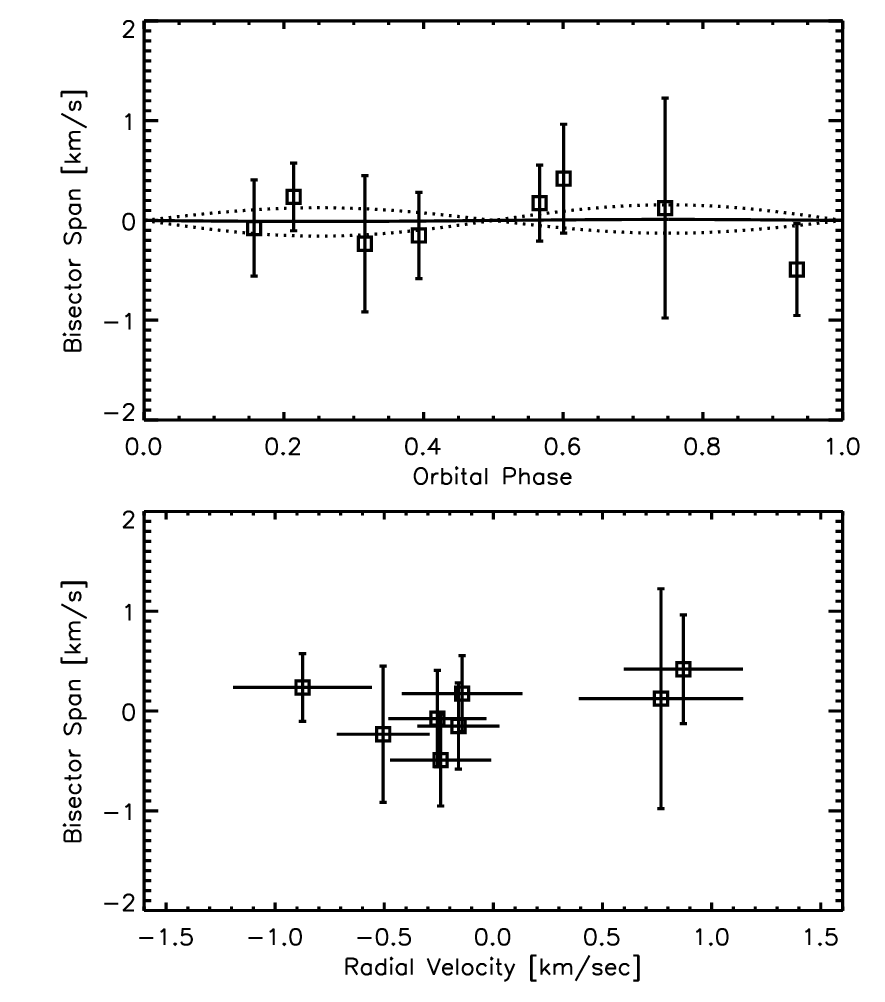,width=0.5\textwidth}
\caption{\label{bisector} Bisector variations as function of orbital phase
(top panel) and radial velocity value (bottom panel). The solid line and 
dashed lines in the top panel indicate the least-squares fitted sinusoidal
variation in the bisector span and its uncertainty at $-0.01\pm0.140$ km s$^{-1}$.}
\end{figure}

We also determined the variations of the bisector span (following Queloz et al. 2001) as function of radial velocity and orbital phase. These are shown 
in figure \ref{bisector}. We least-squares fitted the bisector span measurements as function of orbital phase with a sinusoid, but no significant variations
at a level of $-0.01\pm0.140$ km s$^{-1}$ are found.
Although this means that there is no indication that the measured radial 
velocity variations are due to line shape variations, caused by either stellar 
activity or blends of more than one star, the errors are very large,
making any claim based on the bisector span rather uncertain.

\section{Estimation of the stellar and planetary parameters}

\subsection{Stellar mass, radius, and age}

The transit photometry provides an estimate of the mean density of the 
host star, while the spectroscopic observations yield its surface temperature,
surface gravity, and metallicity. The stellar evolutionary tracks of 
Siess et al. (2000) were subsequently used to estimate the star's mass, radius, 
and age, resulting in M$_{\rm{s}}$=1.52$\pm$0.08 M$_{\rm{sun}}$, 
R$_{\rm{s}}$=1.53$\pm$0.04 R$_{\rm{sun}}$, and an age of $<$0.66 Gyr. 
These parameters correspond to a surface gravity of log g = 4.25$\pm$0.01, 
which is in reasonable agreement, but about 1.7$\sigma$ lower than the 
spectroscopic value. It should be realised that it is notoriously difficult 
to obtain reliable log g values from relatively low signal to noise spectra.

\subsection{Planetary Mass and Radius}

Using the values obtained from the transit fit to the GROND light curves, 
the radial velocity fit, and the stellar parameters as derived above,
we obtain a planetary mass of M$_p$ = 4.5$\pm$1.5 M$_{\rm{jup}}$ and 
a planetary radius of R$_p$ = 1.61$\pm$0.04 R$_{\rm{jup}}$.
The semi-major axis of the orbit is at a=0.0308$\pm$0.0005 AU. The mean density 
of the planet is 1.44$\pm$0.49 g cm$^{-3}$.

This means that OGLE2-TR-L9b is one of the largest known transiting 
hot Jupiters, only TrES-4b and WASP-12b are marginally larger,
although its mean density is similar to that of Jupiter. 
Even so, OGLE2-TR-L9b is significantly larger than expected for 
an irradiated $\sim$4.5 M$_{\rm{Jup}}$ planet (Fressin et al. 2007). 

\section{Rejection of blended eclipsing binary scenarios}

Large photometric transit surveys are prone to produce a significant 
fraction of false interlopers among genuine transiting extrasolar planets.
If the light from a short-period eclipsing stellar binary is blended with
that from a third, brighter star, the combined photometric signal can 
mimic a transiting exoplanet. Although the radial velocity variations 
induced by an eclipsing binary should be orders of magnitude larger
than those caused by a planet, the blending of the spectral lines with those
from the brighter, third star could produce variations in the overal 
cross-correlation profile that have significantly smaller amplitudes,
possibly as small as expected for giant planets. Since this 
would be accompanied with significant line-shape variations, bisector
span analyses are often used to reject a blended eclipsing binary scenario. 

\begin{figure}
\psfig{figure=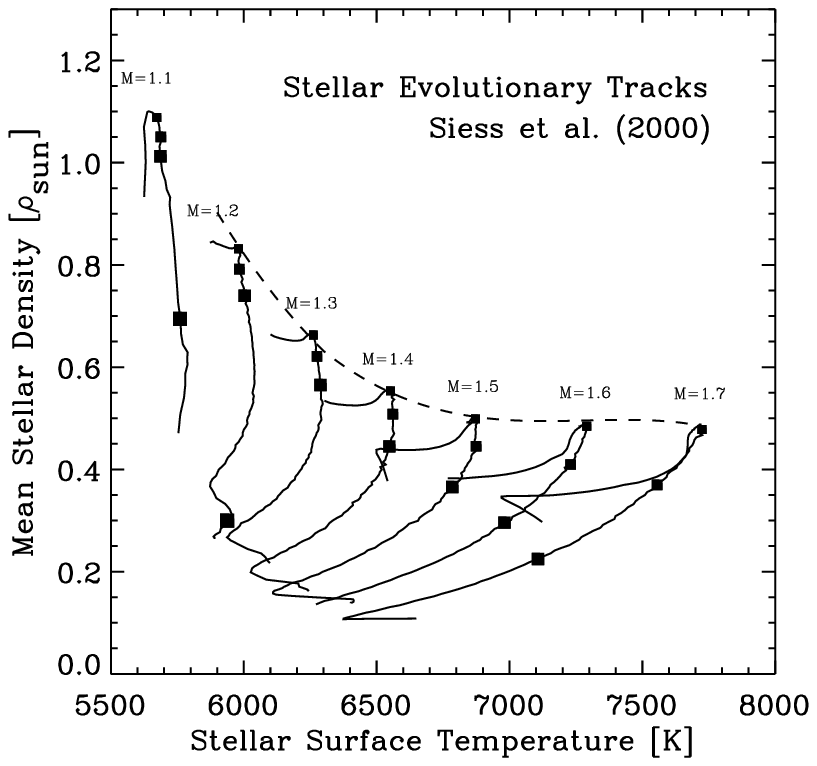,width=8.5cm}
\psfig{figure=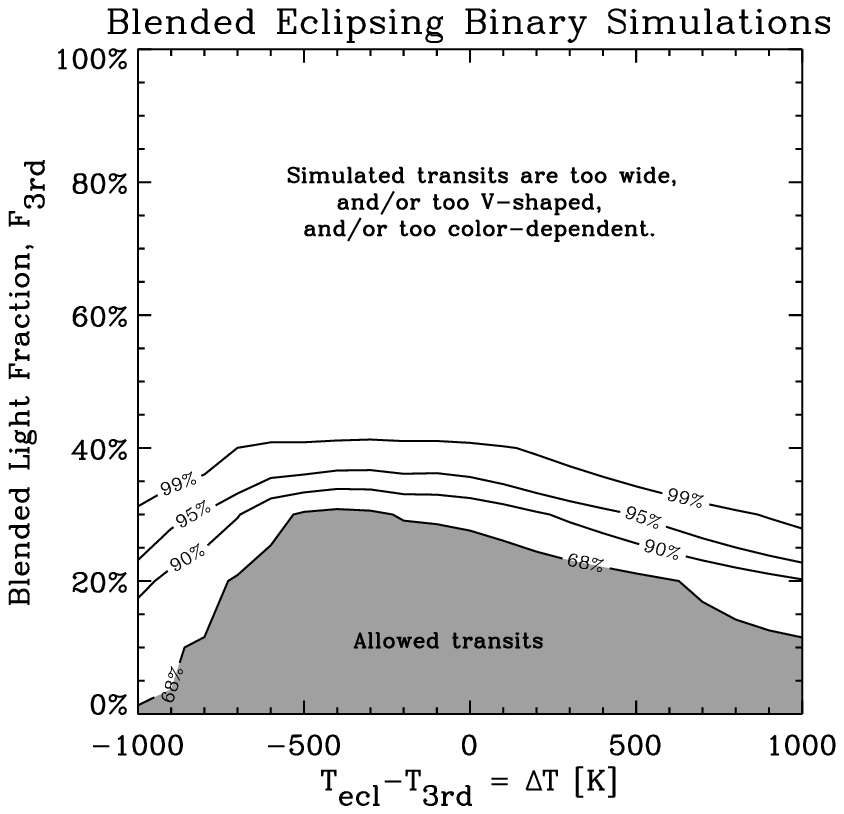,width=8.5cm}
\caption{\label{quant}
The upper panel shows the stellar surface temperature, T$_{\rm{s}}$, versus the 
mean density, $\rho_{\rm{s}}$, for evolutionary tracks of Siess et al. 
(2000). The filled squares on each track indicate stellar ages of 
0.1, 0.5, and 1$\times10^9$ years, (and 5$\times10^9$  years for M$\le$1.2M$_{\rm{Sun}}$), with larger symbols indicated higher ages. 
The dashed line indicates the maximum possible $\rho_{\rm{s}}$ for a given 
stellar surface temperature. 
The bottom panel shows the confidence intervals from the 
$\chi$-square analysis of all possible 
blended eclipsing binary scenarios fitted to the GROND light curves, 
with on the x-axis the difference in surface 
temperature between the eclipsed and the third star,
 and on the y-axis the fraction of the total light 
coming from the third star in r'-band. It shows that the 
combined g'r'i'z' light curves can only be fitted including a low level 
($\lesssim 30\% $) of light contamination.}
\end{figure}

Although no significant variations in the bisector span are observed in 
OGLE2-TR-L9, it could be argued that this is due to the lack of 
signal-to-noise. We show however that a blended eclipsing binary scenario can
be rejected anyway, because of the following observations:
\begin{itemize}
\item[1] \underline{Transit light curves from g$'$ to z$'$ band:}
As can be seen in Fig. 1, there is an excellent agreement between 
the light curves from g to z band. This means that if the transit 
was actually caused by a background eclipsing binary blended with a 
bright foreground star, the colors (and thus the surface temperatures)
of the eclipsed binary star and foreground star should be very similar. 
\item[2] \underline{Transit shape and spectral classification:}
The mean stellar density as determined from the transit photometry is in 
excellent agreement with the spectral classification, both consistent with 
an early F star. Using the argument above, this means that if this is a 
blend, then both the foreground star and the eclipsed binary star should be 
early F stars. 
\end{itemize}
However, if we now assume that a significant fraction of the light comes from 
a foreground star, and we remove this contribution from the light curve, 
the transit can no longer be fitted by an early F star, but only by a star of 
significantly higher mean density, implying a cooler, less massive star,
which is again in contradiction with point 1).
This means that the early F star {\sl is} the transited object, and 
that a blended eclipsing binary scenario can be rejected.
 
To further explore the possible role of additional light from 
a blended star, we performed a quantitative analysis, simulating 
background eclipsing binary systems with their light
diluted by that from a third star. We first used the stellar evolutionary 
tracks of Siess et al. (2000) to determine the full range of stellar 
parameters that can be present in eclipsing binaries, of which only
the stellar surface temperature, T$_{\rm{ecl}}$, and the mean stellar 
density, $\rho_{\rm{ecl}}$ of the eclipsed star are of interest for the 
simulations. Note that the evolutionary status of the third star is not 
important, 
since we do not restrict ourselves to physical triple systems,
but also include chance-alignments of back- and foreground stars. 
As is indicated in the top panel of figure \ref{quant}, where 
the stellar evolutionary tracks are shown, there is a maximum possible 
 mean-stellar density for a given surface temperature. 
This was used as a boundary condition in the simulations. 

In our simulations we varied two parameters, 1) the difference
between the surface temperature of the eclipsed star and that of the third 
star, $\Delta$T ($-$1000 K$<$$\Delta$T$<$+1000 K), and 2) the fraction of light coming 
from the third (possibly unrelated) star, F$_{\rm{3rd}}$ (0$<$$\rm{F}_{\rm{3rd}}$$<$ 99\%).
The combined light of the eclipsing binary and third star 
should produce a spectrum which 
is best fitted with a surface temperature of T$_{\rm{comb}}$ = 6933 K. Therefore a
simple linear relation between  T$_{\rm{comb}}$ and the surface temperatures
of the individual stars was assumed, such that 
$\rm{T}_{\rm{ecl}}=\rm{T}_{\rm{comb}}-\frac{\rm{F}_{\rm{3rd}}}{1+\rm{F}_{\rm{3rd}}}\Delta \rm{T}$. Note that any small fraction of light that could be coming from the eclipsing star is simply added to F$_{\rm{3rd}}$. In this way, 
each combination of F$_{\rm{3rd}}$ and $\Delta$T, results in a T$_{\rm{ecl}}$ and 
a maximum possible $\rho_{\rm{ecl}}$. It also results in a fractional 
contribution of light from the third star that varies over the four filters.

Subsequently, for each combination of F$_{\rm{3rd}}$ and $\Delta$T, model eclipsing binary light curves were least-squares fitted to the g'r'i'z' GROND data, using as before the algorithms of Mandel \& Agol (2002), in which 
the binary size ratio and the impact parameter were completely free to vary, 
and $\rho_{\rm{ecl}}$ was retricted to be below the upper limit set by T$_{\rm{ecl}}$.
In this way, all possible blended eclipsing binary scenarios are 
simulated, independently of whether the third star is physically related to 
the binary or not. The bottom panel of figure \ref{quant} shows the 
confidence contours of the $\chi$-square analysis of all possible blended
eclipsing binary scenarios. It shows that the combined g'r'i'z' data can 
only be fitted by light curves of eclipsing binaries with a low level
$\lesssim$30\% (90\% confidence level) of blended light, meaning that 
most light in the stellar spectrum must come from the eclipsed star. 
Scenarios in which the stellar spectrum is dominated by a third star
with a small contribution from a background eclipsing binary, can 
be strongly rejected. The transit light curves produced by those 
rejected scenarios
are simply too wide, and/or too V-shaped, and/or too color dependent to fit 
the GROND data. One scenario that we cannot reject, is a small contribution
from a blended star. For example, it could in principle be possible that the 
light from the transited F3 star is diluted at a $\sim$30\% level with 
light from another F star (with a similar v$sin$i and radial velocity, 
otherwise it would show up in the spectra). 
In this case, the transiting planet would be 
$\sim$30\% more massive (and $\sim$15\% larger) than determined above, by
no means moving it outside the planet mass range. 

Note that for most transiting
planets presented in the literature, such a low-level contamination 
scenario can not be excluded, since the variations in the bisector span 
would be orders of magnitude smaller than in the case of a blended 
eclipsing stellar binary. This is because the radial velocity variations 
in the latter case are 10$^{2-3}$ times larger than in the first case. 

There have been several reports of blended eclipsing binaries 
hiding out as transiting planets, most notably by Mandushev et al. (2005), and 
Torres et al. (2004). However, these studies dealt with very low 
signal-to-noise light curves, and the true nature of these systems 
would have been easily brought to light 
by such high quality photometric data as presented in this paper. 
Mandushev et al. (2005) rejected a transiting planet scenario 
for the fast rotating (v$sin$i=34 km/s) F5 star GSC~01944-02289, in 
favour of a blended eclipsing binary. This system was shown to  
consist of a hierarchical 
triple composed of an eclipsing binary with G0V and M3V components, in 
orbit around a slightly evolved F5 dwarf. The latter star in this 
scenario contributes for $\sim$89\% to the total light from the system.
Although they claim that the true nature of this system was not revealed 
by their BVI light curves,
the color difference between the G0V and F5V star means that the transit
must be 25-30\% deeper in I-band than in B-band. However, no quantitative
analysis of the light curves was presented, and the authors claim that 
the true nature of the system was only brought to light by spectroscopic 
means. 
In a similar fashion Torres et al. (2004) presented the case of OGLE-TR-33,
which was identified as a triple system consisting of an eclipsing
binary with F4 and K7-M0 components orbiting a slightly evolved F6 star. 
However, their photometry relied solely on the original I-band 
OGLE-III data, resulting in a relatively low SNR transit detection with the 
ingress and bottom of the transit not well covered. They also claim
that the blended eclipsing binary is only revealed by spectroscopy. 
However, their best fitting planet model already pointed towards a
very unlikely planet radius of $\sim$3 R$_{\rm{Jup}}$, and the V-shaped
transit produced by the blended eclipsing binary would have been 
easily picked up by our high precision photometry. Note that while 
Torres et al. (2004) and Mandushev et al. (2005) only consider 
physical triple systems, our analysis presented above includes all
possible scenarios, also those involving chance-alignment of background
or foreground stars.

\section{Discussion}

More than seven years and $>$1000 orbital periods after the last observations of 
OGLE2-TR-L9, the transit signal was rediscovered only 8 minutes from its 
predicted time (from S07). It not only shows that an observing campaign 
with large
time intervals between measurements can produce reliable light curves, 
it also shows it produces extremely accurate orbital periods. 

OGLE2-TR-L9b is the first extrasolar planet discovered transiting a 
fast rotating (v$sin$i=39 km/s) F star. OGLE2-TR-L9 is also the star with 
the highest surface temperature (T=6933 K) of all main sequence stars that
host an exoplanet known to date. 
It is therefore not surprising that the uncertainties 
in the radial velocity variations are higher than for most other transiting 
exoplanets presented in the literature. Only due to the high mass
of OGLE2-TR-L9b, we were able to detect its radial velocity signature. 
Note however that, since a blend scenario can be rejected at high significance, 
an upper limit to the mass of OGLE2-TR-L9b would have been sufficient
to claim the presence of a transiting extrasolar planet, although with an 
unknown mass. Similar arguments
may have to be used in the case of future detection of transits of 
Earth-size planets from Kepler or CoRoT, since their radial velocity 
signature may be too small to measure.

OGLE2-TR-L9b has a significantly larger radius than expected for a 
planet of about 4.5 times the mass of Jupiter, even if it is assumed that 
0.5\% of the incoming stellar luminosity is dissipated at the planet's 
center (Fressin et al. 2007).
However, it is not the only planet found to be too large (e.g. CoRoT-exo-2b, 
TrES-4b, and XO-3b). Several mechanisms have been proposed to explain
these 'bloated' radii, such as more significant core heating
and/or orbital tidal heating (see Liu, Burrows \& Ibgui 2008 for a recent detailed
discussion).

The measured v$sin$i and estimated stellar radius combine to a 
rotation period of the host star of $\sim$1.97$\pm$0.04 days. It 
means that the rotation of the star is not locked to the 
orbital period of OGLE2-TR-L9b. A v$sin$i of 39 km/sec is within
the normal range for stars of this spectral type. 
The mean v{\sl sin}i of  F5 to F0 stars
in the solar neighbourhood range from 10$^{2}$ to 10$^{3}$ km/sec   
respectively. Note that the v{\sl sin}i of OGLE2-TR-L9a is only 
$\sim$9\% of the expected break-up velocity for a star of this 
mass and radius. Assuming the general Roche model for a rotating 
star (e.g. Seidov, 2004), the ratio of polar to equatorial radius
of OGLE2-TR-L9a will be on the order of, 
$1-\frac{1}{2}(v/v_{\rm{max}})^2\sim$~0.996. Thus, the rotational
flattening of the host star is not expected to significantly 
influence the transit shape. 
OGLE2-TR-L9 is expected to exhibit a strong Rossiter-McLaughlin effect.
Simulation using a segmented stellar surface predict an amplitude of 
230 m/sec.

\begin{acknowledgements}
We thank the anonymous referee very much for his or her insightful comments.
Based on observations collected at the European Organisation for Astronomical Research in the Southern Hemisphere, Chile (280.C-5036(A))
T.K. acknowledges support by the DFG cluster of excellence 'Origin and
Structure of the Universe'.
\end{acknowledgements}

\end{document}